\documentclass[]{article}
\usepackage{lmodern}
\usepackage{amssymb,amsmath}
\usepackage{ifxetex,ifluatex}
\usepackage{fixltx2e} 
\ifnum 0\ifxetex 1\fi\ifluatex 1\fi=0 
  \usepackage[T1]{fontenc}
  \usepackage[utf8]{inputenc}
\else 
  \ifxetex
    \usepackage{mathspec}
  \else
    \usepackage{fontspec}
  \fi
  \defaultfontfeatures{Ligatures=TeX,Scale=MatchLowercase}
\fi
\IfFileExists{upquote.sty}{\usepackage{upquote}}{}
\IfFileExists{microtype.sty}{%
\usepackage{microtype}
\UseMicrotypeSet[protrusion]{basicmath} 
}{}
\usepackage[margin=1in]{geometry}
\usepackage{hyperref}
\hypersetup{unicode=true,
            pdftitle={Improving information quality of Wikipedia articles with cooperative principle},
            pdfauthor={Miloš Fidler and Dejan Lavbič},
            pdfborder={0 0 0},
            breaklinks=true}
\usepackage{natbib}
\usepackage{longtable,booktabs}
\usepackage{graphicx,grffile}
\makeatletter
\def\maxwidth{\ifdim\Gin@nat@width>\linewidth\linewidth\else\Gin@nat@width\fi}
\def\maxheight{\ifdim\Gin@nat@height>\textheight\textheight\else\Gin@nat@height\fi}
\makeatother
\setkeys{Gin}{width=\maxwidth,height=\maxheight,keepaspectratio}
\IfFileExists{parskip.sty}{%
\usepackage{parskip}
}{
\setlength{\parindent}{0pt}
\setlength{\parskip}{6pt plus 2pt minus 1pt}
}
\ifx\paragraph\undefined\else
\let\oldparagraph\paragraph
\renewcommand{\paragraph}[1]{\oldparagraph{#1}\mbox{}}
\fi
\ifx\subparagraph\undefined\else
\let\oldsubparagraph\subparagraph
\renewcommand{\subparagraph}[1]{\oldsubparagraph{#1}\mbox{}}
\fi

\let\rmarkdownfootnote\footnote%
\def\footnote{\protect\rmarkdownfootnote}

\usepackage{titling}

  \title{Improving information quality of Wikipedia articles with cooperative
principle}
    \pretitle{\vspace{\droptitle}\centering\huge}
  \posttitle{\par}
    \author{Miloš Fidler and Dejan Lavbič}
    \preauthor{\centering\large\emph}
  \postauthor{\par}
    \date{}
    \predate{}\postdate{}

\usepackage{booktabs}
\usepackage{longtable}
\usepackage{array}
\usepackage{multirow}
\usepackage[table]{xcolor}
\usepackage{wrapfig}
\usepackage{float}
\usepackage{colortbl}
\usepackage{pdflscape}
\usepackage{tabu}
\usepackage{threeparttable}
\usepackage{threeparttablex}
\usepackage[normalem]{ulem}
\usepackage{makecell}

\usepackage{amsthm}

\theoremstyle{definition}

\theoremstyle{definition}

\theoremstyle{definition}

\theoremstyle{remark}

\begin{document}
\maketitle

\begin{quote}
Miloš Fidler and \textbf{Dejan Lavbič}. 2018.
\href{https://doi.org/10.1108/OIR-01-2016-0003}{\textbf{``Improving
information quality of Wikipedia articles with cooperative
principle''}}, \href{https://www.emeraldinsight.com/loi/oir}{Online
Information Review \textbf{(OIR)}}, 41(6), pp.~797 - 811.
\end{quote}

\section*{Abstract}\label{abstract}
\addcontentsline{toc}{section}{Abstract}

\textbf{Purpose} -- The purpose of this paper is to investigate the
impact of cooperative principle on the information quality (IQ) by
making objects more relevant for consumer needs, in particular case
Wikipedia articles for students.

\textbf{Design/methodology/approach} -- The authors performed a
quantitative study with participants being invited to complete an online
survey. Each rater evaluated three selected and re-written articles from
Wikipedia by four IQ dimensions (accuracy, completeness, objectivity,
and representation). Grice's maxims and submaxims were used to re-write
articles and make them more relevant for student cognitive needs. The
results were analyzed with statistical methods of mean, standard
deviation, Cronbach's \(\alpha\), and ICC (two-way random model of
single measure).

\textbf{Findings} -- The study demonstrates that Wikipedia articles can
be made more relevant for student needs by using cooperative principle
with increase in IQ and also achieving higher consistency of students'
scores as recent research. In particular, students in the research
perceived the abstract, constructed with cooperative principle, more
objective and complete as reported in recent research.

\textbf{Practical implications} -- The work can benefit encyclopedia
editors to improve IQ of existing articles as well as consumers that
would obtain more relevant information in less reading time.

\textbf{Originality/value} -- This is one of the first attempts to
empirically investigate the application of cooperate principle to make
objects more relevant for consumer needs and impact of this on IQ. IQ
improvement evidence is provided and impacts on IQ dimensions such as
objectivity, completeness, accuracy, and representation for research
community to validate and compare results.

\section*{Keywords}\label{keywords}
\addcontentsline{toc}{section}{Keywords}

Conversation maxims, Cooperative principle, Improving information
quality, Information quality assessment and analysis, Interrater
reliability, Presenting information in relevant way

\section{Introduction}\label{introduction}

The international Data Corporation reported that the total amount of
global data surpassed \(1.8\) zettabyte in \(2011\) and it is predicted
to reach \(35\) zettabyte by \(2020\) \citep{kambatla_trends_2014}.
Economies, companies, and our daily activities are becoming more and
more data driven \citep{zhang_domain_2005, fidler_research_2015}. As a
result, demand for high-quality information is increasing; however, we
are still struggling to understand how quality of information can be
improved. Consumers are not interested in just any information; they
request the best information available for their purpose
\citep{mai_quality_2013}. This elusive trait of information, how well it
serves consumers' needs or how fit it is for use, was investigated
further by \citep{wang_beyond_1996} into definition of information
quality (IQ) as a multi- dimensional concept with dimensions such as
accuracy, consistency, completeness, timeliness, and representation.

The majority of the previous IQ research was conducted by employing
students to measure quality of online objects, mostly Wikipedia articles
\citep{kogut_what_1996, stenmark_leveraging_2000}. Authors investigated
how students perceived IQ dimensions, but did not advance their research
with recommendations regarding IQ improvement
\citep{metzger_making_2007}. A step forward was made by
\citep{mai_quality_2013}, who theoretically discussed the possibility of
using the Grice's maxims to construct objects from which consumers could
retrieve more relevant information for specific purpose or context of
use. Grice's maxims are recommendations used within the cooperative
principle for effective communication between two conversing parties,
such as be relevant and clear, make messages as informative as possible,
and communicate only true and confirmable facts
\citep{grice_logic_1967}. This research aims to experimentally test
whether Gricean principle can make Wikipedia articles fit for students'
needs, thus improve IQ.

The contribution of this study is to empirically and statistically prove
that Gricean principle, if applied to customize information in objects
for specific consumers' use, can improve IQ. This study provides
evidence of IQ improvement when Wikipedia content is made fit for
students' needs with an approach based on Gricean principle. Research
community is encouraged to conduct similar IQ improvement studies, not
only for students but also for other groups (e.g.~young professionals,
retired people, etc.). Furthermore, results can be beneficial for
Wikipedia and other encyclopedias since they can apply Gricean principle
in similar way to improve their content for students' needs.

Additionally, the paper statistically analyzes and interprets students'
evaluations of Wikipedia articles by dimensions of accuracy,
completeness, representation, and objectivity. This provides the
research community with informed insights on students' perception of IQ,
which can be used as a reference point in future studies. IQ researchers
who employ students as evaluators can validate their results, whereas
non-students-based evaluation studies can compare and further generalize
their findings.

The remainder of the paper is organized as follows. Section \ref{iq}
presents a comprehensive review of related work about IQ evaluation and
improvement. Section \ref{method} bestows research method for estimating
impacts on IQ when cooperative principle is used to make Wikipedia
articles more relevant for students' needs. Section \ref{results}
reports on the results of students' perception of quality with
discussion and implication. The paper concludes with Section
\ref{conclusion}, where reflections on the findings are included and
possible directions for future research are discussed.

\section{IQ}\label{iq}

\subsection{Construct taxonomy}\label{construct-taxonomy}

There is no agreed-upon definition of IQ \citep{michnik_assessment_2009}
and despite significant interest in IQ the domain still remains quite
immature \citep{baskarada_critical_2014}. Especially, because the world
``quality'' characterizes non-physical construct, namely information,
which when retrieved and consumed can have different meaning for
different users. While some argue that data has only meaning if put into
context, thus becoming information \citep{vrhovec_outsourcing_2015},
others emphasize objective and subjective view on information.

Objective view on information is defined as observer and situation
independent \citep{hjorland_information:_2007} and clarified in
\citep{bates_introduction_2005} that any (observer and situation
independent) difference produces information and therefore also is
information. Based on this view, IQ is discussed from external aspect
\citep{arazy_measurability_2011} as the degree to which information
meets specified and generally accepted requirements
\citep{eppler_managing_2006}. These are usually quality specifications
\citep{ge_review_2007} set for optimal data values stored in a database
\citep{savchenko_automating_2003} to avoid deficiencies between the real
world state and information system representation
\citep{wand_anchoring_1996}.

\citep{hjorland_information:_2007} argues that subjective definition of
information is at least as or even more important. Subjective aspect
(also pointed out as subjectivity) is explained as observer and
situation dependent and illustrated in
\citep{hjorland_information:_2007} as a difference that makes a
difference (for somebody or for something or from a point of view). From
this perspective IQ is considered as individual's ``subjective judgment
of goodness and usefulness'' \citep{hilligoss_developing_2008} or the
degree to which the information meets the expectations of the user
\citep{eppler_managing_2006}.

Wang and Strong (1996) adopted ``fitness for use'' definition, which
considers the consumer's perspective, embodying both the objective and
subjective perspective of consumed information \citep{wang_beyond_1996}.
Using a two-stage survey, they developed a framework with four
categories and 15 dimensions which are generally accepted in the
literature \citep{baskarada_critical_2014}. Additionally, other
researchers developed a range of IQ evaluation metrics
\citep{madhikermi_data_2016}, assessment instruments
\citep{suhardi_total_2014} and frameworks
\citep{fink-shamit_information_2008, hilligoss_developing_2008, rieh_credibility:_2007, bates_information_2005},
which can be used to assess different aspects of IQ.
\citep{knight_developing_2005} analyzed characteristics of 12 such IQ
frameworks, including Wang and Strong's and concluded that each
framework is based upon an author's viewpoint, meaning that IQ dimension
are combined in different ways \citep{ge_review_2007}, such as
hierarchical \citep{wang_beyond_1996}, ontological
\citep{wand_anchoring_1996}, semiotic \citep{helfert_managing_2001},
product and service \citep{kahn_information_2002}.

Recent research emphasized the importance of IQ within wide range of
industries, including online communities
\citep{mohammadi_is_2015, font_assessing_2015, zheng_impacts_2013, detlor_information_2013},
financial industry \citep{lee_enhancing_2016, corona_accounting_2015},
healthcare \citep{lopez_information_2016, ceylan_information_2016},
digital media \citep{romero-rodriguez_dimensions_2016}, and tourism
\citep{berezan_impact_2016, paglieri_trusting_2014}, and demonstrated
the business impact that retrieved high-quality information can have on
supply chain \citep{zhou_supply_2014}, risk management
\citep{corona_accounting_2015, nicolaou_information_2013}, reporting
\citep{madhikermi_data_2016}, innovativeness, and stock market return
\citep{lee_enhancing_2016}, as well as other impacts, such as positive
consumer trust \citep{berezan_impact_2016}, user website satisfaction
\citep{bastida_performance_2014}, perceived website quality, trust and
usefulness \citep{ghasemaghaei_macro_2016, leite_model_2016}, user's
decision making \citep{petter_information_2013, shen_unleash_2013},
customer loyalty, blogging success \citep{wang_model_2014}, and
re-purchase \citep{ghasemaghaei_macro_2016}. There have been, however,
few examinations and validated suggestion on how to improve IQ of
objects, due to the opposing perceptions of quality among information
consumers.

\subsection{Students, Wikipedia use and
assessment}\label{students-wikipedia-use-and-assessment}

In recent years Wikipedia received a great interest from research
community with the attention on how good is the quality of its articles.
Several authors
\citep{denning_wikipedia_2005, luyt_improving_2008, wallace_democratization_2005}
expressed concerns about the quality of Wikipedia as a source and
interest in evaluation of its content. Overall, people perceived the
quality of Wikipedia articles as ``quite good''
\citep{chesney_empirical_2006, stvilia_information_2008} and often read
the articles to obtain additional knowledge \citep{fallis_toward_2008}.
Several other studies indicated
\citep{brown_wikipedia_2011, clauson_scope_2008} that Wikipedia users
should worry more that Wikipedia articles are incomplete and inaccurate.

There were many initiatives that information-seeking public should be
examining and controlling the quality
\citep{ghasemaghaei_macro_2016, zheng_impacts_2013}. As a result, many
IQ studies are performed with Wikipedia articles, whereas students are
employed as IQ assessors \citep{mesgari_sum_2015}. Despite controversies
with student citation of Wikipedia and concerns about poor gatekeeping
(e.g.~editorial or peer review)
\citep{arazy_measurability_2011, helfert_impact_2013}, many researchers
do in fact cite Wikipedia and even promote its use
\citep{okoli_wikipedia_2014}.

Wikipedia is open for everyone, easily accessible, and interactive
\citep{mai_quality_2013}, and it provides a unique opportunity for
educating students in digital literacy \citep{okoli_wikipedia_2014}. In
a recent literature review, \citep{okoli_wikipedia_2014} reported 34 IQ
studies in which researches investigated how students use Wikipedia as a
general source of information and how they were assigned work that
explicitly involved reading Wikipedia articles. \citep{lim_gender_2010}
compared student usage of Wikipedia by gender. They found that while
male students used Wikipedia more frequently and had a positive attitude
toward it, female students displayed more cautious or conservative
attitudes, emotions, and behaviors.

Students like Wikipedia, since it is very comprehensive and easily
readable, but at the same time they are well aware of its limitation and
they make the best of it \citep{korosec_chemical_2010}.
\citep{shaw_wikipedia_2008} indicated that most students used Wikipedia
when they were unfamiliar with the topic as a starting point to obtain
basic knowledge that lead to other sources. When students were familiar
with the topic, they used Wikipedia to gather additional information
about the topic. \citep{okoli_wikipedia_2014} identified two use cases:
personal (source of information) or academic (citations) use. Even when
students use Wikipedia for academic purpose they are well aware of its
limitations exploiting Wikipedia to find more ``reputable'' sources.
\citep{choolhun_google:_2009} for instance documented that Wikipedia is
increasingly being used as the first source for legal information
inquiries by law students. \citep{waters_why_2007} recommended Wikipedia
to students by saying that Wikipedia is a fine place to search for a
paper topic or begin the research process. \citep{patch_meeting_2010}
concluded that by employing Wikipedia, students can have an easier time
making the leap to higher-level inquiry and responsible scholarship.

\subsection{Improving IQ}\label{improving-iq}

Literature review showed that there is a gap regarding IQ improvement
studies. Most of the assessment studies are descriptive, further
clarifying taxonomy of IQ by proposing new dimensions. However, they
fail to provide actions for IQ improvement of Wikipedia articles.
Although it was identified which dimensions assessors consider the most
relevant at IQ evaluation, it was at the same time noted that assessors
interpret and use dimension values differently.
\citep{arazy_measurability_2011} revealed the full extent of this issue,
empirically illustrating why past research was not able to identify IQ
improvement measures. IQ researchers are therefore challenged to find
alternative approaches to identify IQ improvement measures and prove the
significance of their impact.

Although cooperative principle with conversation maxims arise from
pragmatics of natural language and are used to improve conversational
effectiveness of communication, \citep{mai_quality_2013} proposed and
discussed their use in the field of IQ. The author elaborated that IQ
must be understood in a context in which the consumer is situated while
retrieving information. Information producers should therefore act as if
they are actually speaking with information consumers and overcome the
disordered nature of a language by applying Gricean principle
\citep{grice_logic_1967} for successful communication referred to as
maxims and submaxims.

Studied literature review showed that there are no empirical studies,
which investigated the application of cooperative principle to improve
IQ. It is assumed that presenting information in a more relevant way for
the specific use improves the perception of IQ, but it is yet not known
to what extent and by what rate of agreement among consumers. Thus,
knowing why users' need information provides one with the capability for
questioning poor IQ and suggesting new solutions for its improvement.

This research narrows focus on investigation of IQ improvement of
Wikipedia articles when Gricean principle is applied. Therefore, main
research question is as follows:

\begin{quote}
\textbf{RQ1}. Can Grice's maxims and submaxims be applied to Wikipedia
articles to improve their quality for students?
\end{quote}

In particular, this paper investigates what the overall IQ affect rate
is and whether there are some recognized IQ dimensions that are more
affected than others.

\section{Method}\label{method}

To test the effect of cooperative principle on Wikipedia articles, a
quantitative study with \(265\) students of computer information studies
at entry level (\(35\) percent female, \(65\) percent male; from \(20\)
to \(26\) years of age, with the mean of \(21,90\)) was performed. Each
student was invited to complete an online survey active from January
2015 until December 2017, where they were asked to evaluate three
selected and re-written articles from Wikipedia, presented to them in a
random order by four IQ dimensions: accuracy, completeness, objectivity,
and representation as presented in Table \ref{tab:IQ-dimensions}.
Selected IQ dimensions represented most relevant top-level IQ categories
(intrinsic, contextual, and representational) and were comparable to
dimensions used in other research
\citep{sackmann_cultural_1991, arazy_measurability_2011}.

\begin{table}

\caption{\label{tab:IQ-dimensions}Selected IQ dimensions}
\centering
\begin{tabular}[t]{ll}
\toprule
Dimension & Description\\
\midrule
Accuracy & Information in the article is accurate.\\
Completeness & The article is complete and includes all necessary information.\\
Objectivity & The article is objective; it represents objective opinion about presented topic.\\
Representation & The article is presented consistently and formatted concisely.\\
\bottomrule
\end{tabular}
\end{table}

In January 2015 three original Wikipedia articles with different topics:
human (homo-sapiens from ``People and self'' Wikipedia category),
alexandrite (gemstone with color changing ability from ``Math science
and technology'' Wikipedia category) and Occitan language (Romance
language mostly spoken in Southern France from ``Culture and arts''
Wikipedia category) were selected. Participants had extensive domain
knowledge about human article opposed to narrow domain knowledge about
the alexandrite and Occitan language article. Articles from selected
topics established similar domain knowledge distribution across
participants' as in recent study \citep{arazy_measurability_2011}.

Grice's maxims and submaxims were used to re-write original Wikipedia
articles and make them more relevant for student cognitive needs,
identified and described in \citep{lim_how_2009}. Gricean principle was
applied in such a manner that information was presented to participants
in the same, standardized way, serving both needs, balancing general
with detailed information. This avoided possible user preference toward
characteristics of object which was used to present the information,
e.g.~the length of an article abstract, the number of included sources,
the writing style, and the reputation of included references
\citep{knight_developing_2005}.

In the online survey \(83\) percent of students replied to be annoyed by
low IQ while \(16\) percent were not annoyed by it. They were asked to
evaluate each re-written Wikipedia article by all four IQ dimensions
with Likert scale of \(1\) (strongly disagree) to \(7\) (strongly
agree). All three articles were presented to students in a random order
as well as the order of evaluated dimensions was randomized per article
to avoid confounding bias \citep{pannucci_identifying_2010}. Collected
assessments were analyzed with statistical methods: mean, standard
deviation, Cronbach's \(\alpha\), and ICC (two-way random model of
single measures). In order to answer our research questions, we compared
the results of our student subgroup with the recent study of
\citep{arazy_measurability_2011} and discussed results.

\subsection{Proposed approach of applying Gricean
principle}\label{proposed-approach-of-applying-gricean-principle}

Cooperative principle was used to shorten original Wikipedia articles
\citep{mai_quality_2013} and made them more relevant for typical
Wikipedia users, namely students. \citep{lim_how_2009} reported that
students use Wikipedia to satisfy their cognitive needs, such as: look
for quick facts, to learn something that they are not familiar with and
to get more information about familiar topics. Therefore, original
Wikipedia articles, obtained on January 2015, were summarized and
structured in paragraphs combining general information about most
important entity facts in detail, according to Grice's principles of
conversation maxims and submaxims from Table \ref{tab:Grice-maxims}.

\begin{table}

\caption{\label{tab:Grice-maxims}Grice’s conversation maxims and submaxims used to re-write Wikipedia articles}
\centering
\begin{tabu} to \linewidth {>{\raggedright\arraybackslash}p{2cm}>{\raggedright}X>{\raggedright\arraybackslash}p{7cm}}
\toprule
Category & Maxims and submaxims & Applied action\\
\midrule
C. Relation & C1. Be relevant & Articles summaries were structured with consecutive 5 paragraphs: definition (explained object, discussed in article), differentiation (explained differences from similar or related objects), history (explained origin, historic development and current status of the object), basic characteristics (explained characteristics of the objects such as average size, color, quantity, etc.) and social impact (explained impact on society).\\
\cmidrule{1-3}
 & A1. Make your contribution as informative as is required (for the current purposes of exchange) & Article summaries were made as informative as possible within structured paragraphs.\\
\cmidrule{2-3}
\multirow{-2}{2cm}{\raggedright\arraybackslash A. Quantity} & A2. Do not make your contribution more informative than is required & Two pictures were included in the article.\\
\cmidrule{1-3}
 & B1. Do not say what you believe to be false & \\
\cmidrule{2-2}
\multirow{-2}{2cm}{\raggedright\arraybackslash B. Quality} & B2. Do not say that for which you lack adequate evidence & \multirow{-2}{7cm}{\raggedright\arraybackslash Summarized information was cross-checked by various other sources, such as encyclopedias (Britannica, Scholarpedia, Encarta, etc.) and specialized sites for chosen topics.}\\
\cmidrule{1-3}
 & D1. Avoid obscurity of expression & \\
\cmidrule{2-2}
 & D2. Avoid ambiguity & \\
\cmidrule{2-2}
 & D3. Be brief (avoid unnecessary prolixity) & \\
\cmidrule{2-2}
\multirow{-4}{2cm}{\raggedright\arraybackslash D. Manner (be perspicuous)} & D4. Be orderly & \multirow{-4}{7cm}{\raggedright\arraybackslash Concise and orderly writing style was used.}\\
\bottomrule
\end{tabu}
\end{table}

Articles summaries were structured with five consecutive paragraphs,
each of two to three sentences long. In the first paragraph entity used
in the article was described. Next, it was explained in more details how
this particular entity differs from similar or related entities. Then
two paragraphs were used to present basic information about entity's
history (explained origin, historic development and status of the
entity) and to explain basic characteristics of the entity, such as
average size, color, quantity, etc. In the last two paragraphs, the
basic information about entities' impact on society and a couple of
interesting, specific facts (trivia) associated with entity were added.

\begin{figure}

{\centering \includegraphics[width=1\linewidth]{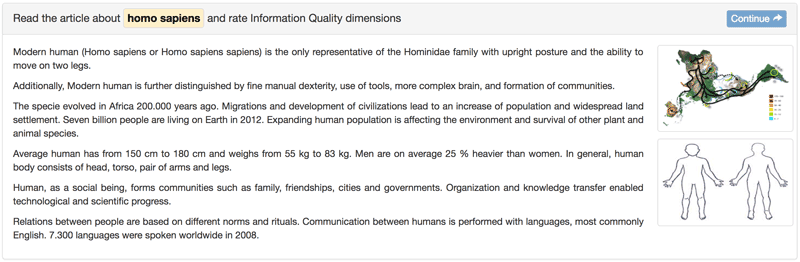}

}

\caption{Outlook of re-written human Wikipedia article with applied Gricean principle}\label{fig:article}
\end{figure}

In total, each re-written article had four paragraphs (\(1\), \(3\),
\(4\), \(5\)) and a picture (how entity looks) to present general
information about the entity and two paragraphs (\(2\) and \(6\)) and a
picture to present more detailed information. Reference pages,
quotations or other sources used in the original Wikipedia article were
skipped to avoid potential biases or preferences of participants toward
the, e.g., credibility of some sources. Final objects, re-written
Wikipedia article, used in the research had title, summary (\(1\) -
\(6\) paragraphs), and two pictures as shown in Figure
\ref{fig:article}. This approach of object construction significantly
reduces the amount of used words in re-written articles compared to
Wikipedia originals. For instance, the article about Human when newly
constructed had only \(180\) words compared to \(12.000\) words of the
same original article on Wikipedia.

\subsection{Data analysis}\label{data-analysis}

Information quality \(IQ^{(dim)}\) for a given dimension \(dim\)
(accuracy, completeness, representation, and objectivity) was defined as
a mean value of all marks given by \(k\) \emph{raters} and is given as
follows:

\[
IQ^{(dim)} = \frac{1}{k}\sum_{i=1}^{k} x_i^{(dim)}
\] where \(x_i\) refers to a specific rater score.

In order to calculate agreement level for all four IQ dimensions,
interclass correlation (\(ICC(2, 1)\), agreement in two-way model with
single measures) was employed and defined as:

\[
\frac{var(\beta)}{var(\alpha) + var(\beta) + var(\epsilon)}
\]

where \(var(\beta) = \frac{BMS - WMS}{k}\) is a variability due to
differences in the objects, \(var(\alpha) = \frac{JMS - EMS}{n}\) is a
variability due to differences in ratings levels used by raters and
\(var(\epsilon) = EMS\) is a variability due to differences in the
evaluations of the objects by the raters. \(BMS\) is the subject mean
square, \(WMS\) is the residual mean square, \(JMS\) is the rater mean
square, and \(EMS\) is the object/rater mean square.

When calculating agreement level for specific IQ dimensions normalized
standard deviation
\(AL_{nsd}^{(dim)} = 1 - \frac{s^{(dim)}}{s_{max}^{(dim)}}\) was used,
where standard deviation \(s^{(dim)}\) is defined as
\(\sqrt{\frac{(x_i^{(dim)} - \bar{x^{(dim)}})^2}{n - 1}}\) and maximum
standard deviation \(s_{max}^{(dim)}\) on a seven-point Likert scale is
defined as \(3 \cdot \sqrt{\frac{n}{n - 1}}\). A similar approach for
calculating agreement level was employed in
\citep{haakonsen_dahl_intra-_2014, moe-nilssen_criteria_2008}, where
authors relied on absolute reliability and the smallest detectable
difference.

\section{Results and discussion}\label{results}

In order to validate and analytically investigate if Wikipedia articles,
which were made fit for students with applied Gricean principle, had
improved IQ, this paper's results \textbf{(FLR)} were compared with
Arazy and Kopak research \textbf{(AKR)}
\citep{arazy_measurability_2011}. Arazy and Kopak used the same IQ
dimensions, scale, and user group (student in early twenties) to measure
IQ. In Section \ref{evidence-of-IQ-improvement} evidence of IQ
improvement is supported with results and three key findings. Results
about student's perception of individual IQ dimensions are presented in
Section \ref{results}, followed by a discussion of the obtained findings
in Section \ref{discussion}.

\subsection{Evidence of IQ
improvement}\label{evidence-of-IQ-improvement}

Results of article evaluations are presented as key statistical IQ
indicators in Table \ref{tab:research-results} and compared with results
published in the AKR. The average value of scores for all four IQ
dimensions (CIQ) increased by \(0,27\) compared to the AKR. At the same
time reliability of construct was improved by \(0,18\) (Cronbach's
\(\alpha\)) and consistency among raters increased by \(0,03\)
(\(ICC(2,1)\)). The values of Cronbach's \(\alpha\) were well above the
required \(0,75\) threshold \citep{straub_validation_2004} and
\(ICC(2,1)\) had a \emph{p-value} lower than the \(5\%\) significance
level. These indicators confirmed that students in the FLR sample
perceived IQ of re-written Wikipedia articles to be higher than students
who evaluated IQ of original Wikipedia articles in the AKR.

\begin{table}

\caption{\label{tab:research-results}Overall research results compared to AKR}
\centering
\begin{tabular}[t]{lrrr}
\toprule
Indicator & FLR & AKR & diff\\
\midrule
Number of students & $265,00$ & $270,00$ & $-5,00$\\
Reliability Cronbach's $\alpha$ & $0,98$ & $0,80$ & $0,18$\\
Consistency $ICC(2,1)$ & $0,20$ & $0,17$ & $0,03$\\
CIQ mean value & $5,16$ & $4,89$ & $0,27$\\
\bottomrule
\end{tabular}
\end{table}

Additionally, difference in perception of CIQ was also investigated
internally, only for students who participated in FLR research, by
dividing them in two groups: those who answered to be annoyed by low IQ
(\(83\%\)) and those who were not (\(17\%\)).

\begin{figure}

{\centering \includegraphics[width=0.4\linewidth]{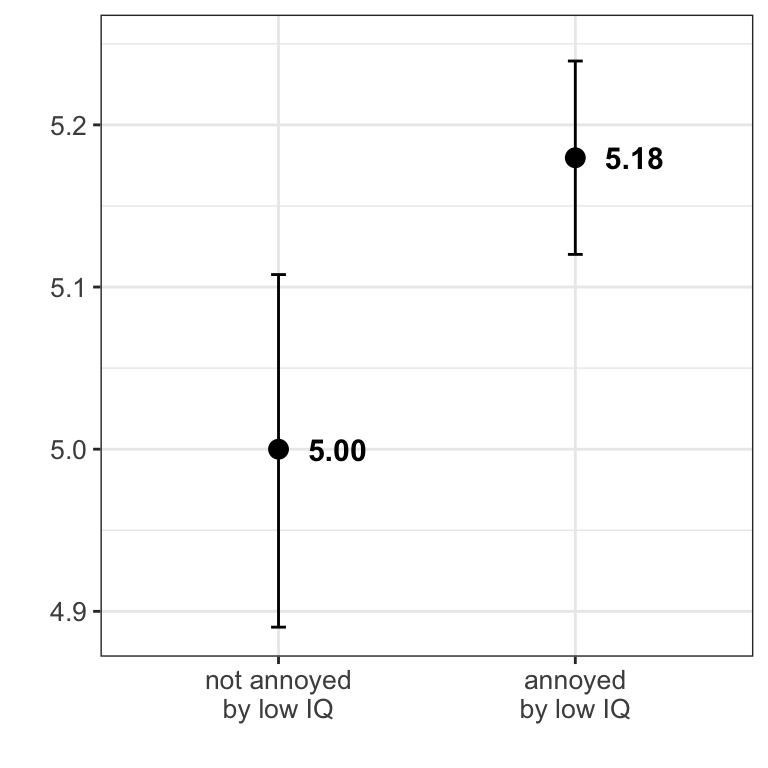}

}

\caption{CIQ means with confidence intervals for students who answered not to be annoyed by low IQ and those who were annoyed}\label{fig:annoyed-by-low-IQ}
\end{figure}

Difference in evaluation of CIQ between both groups
\(\overline{IQ_{\text{not annoyed by low IQ}}^{(CIQ)}} = 5,00 < \overline{IQ_{\text{annoyed by low IQ}}^{(CIQ)}} = 5,18\)
is presented in Figure \ref{fig:annoyed-by-low-IQ} and is statistically
significant (Mann-Whitney-Wilcoxon test; \(\alpha = 0,05\)) with
\(W = 6,32 \times 10^{-5}\) and \(p = 0,0153\). This indicates that
students who considered themselves more sensitive to low IQ recognized
application of Gricean principle as an improvement. Furthermore, even
students who were not annoyed by low IQ considered Wikipedia article
re-written with Gricean principle still to have higher CIQ (\(5,00\))
than original Wikipedia article in the AKR (\(4,89\)).

\subsection{Results of IQ dimensions evaluations}\label{results}

\begin{figure}

{\centering \includegraphics[width=0.5\linewidth]{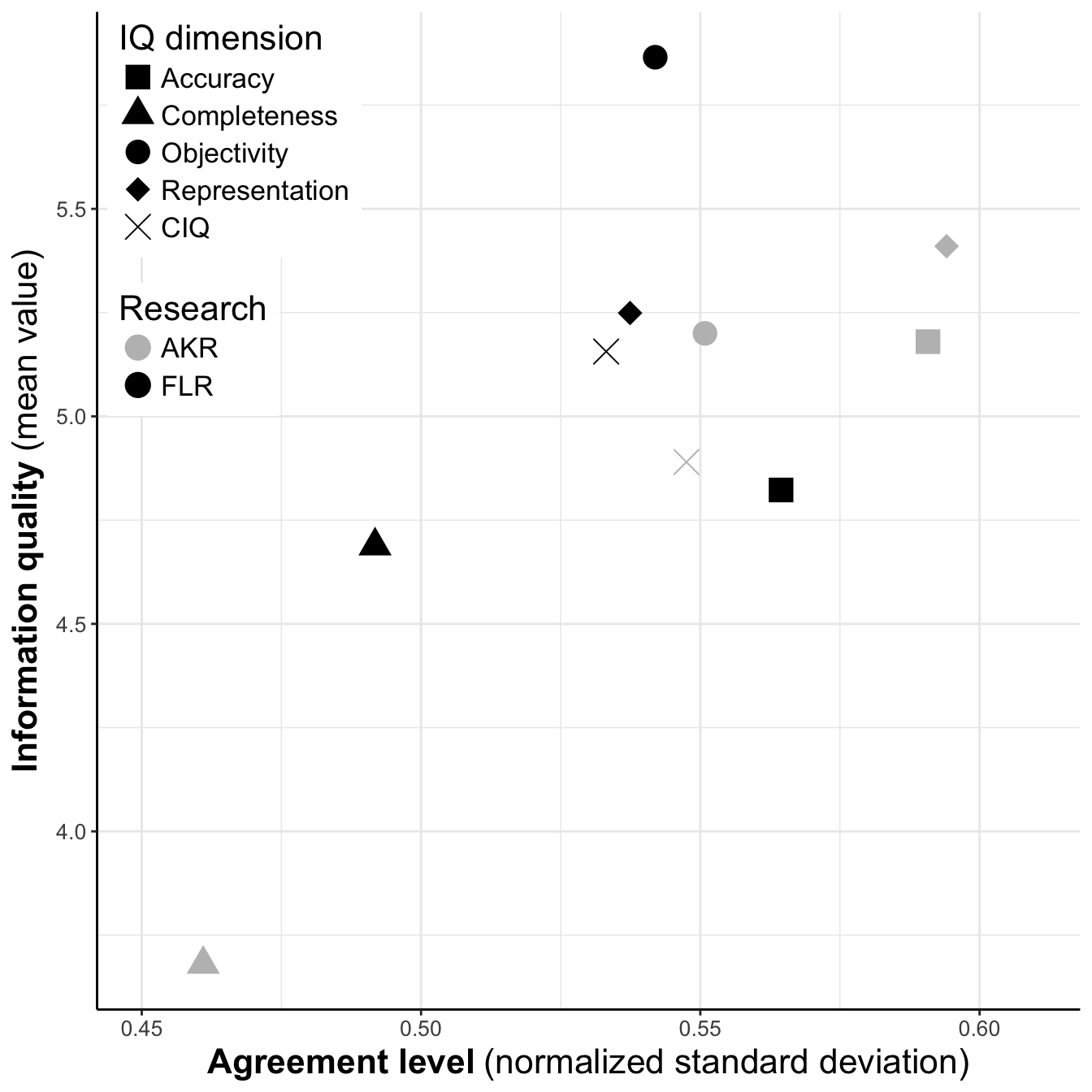}

}

\caption{Impact of individual IQ dimensions on IQ and agreement, compared to AKR}\label{fig:IQ-by-dim}
\end{figure}

Impact on quality (mean value of scores) and agreement level (normalized
standard deviation of scores) of making Wikipedia content fit for
students' needs is further presented by each IQ dimension in Figure
\ref{fig:IQ-by-dim}. AKR reported accuracy, objectivity, and
representation to be similar in size regarding absolute mean and
agreement level values. Completeness on the other hand had low values of
both indicators. Making Wikipedia articles more relevant to students'
needs changed the proportions of these measures. In terms of IQ mean
value three levels were formed: high, which was achieved by objectivity;
moderate, obtained by representation; and low, achieved by accuracy and
completeness. The agreement level of accuracy, completeness, and
representation was higher as of completeness.

\begin{figure}

{\centering \includegraphics[width=0.5\linewidth]{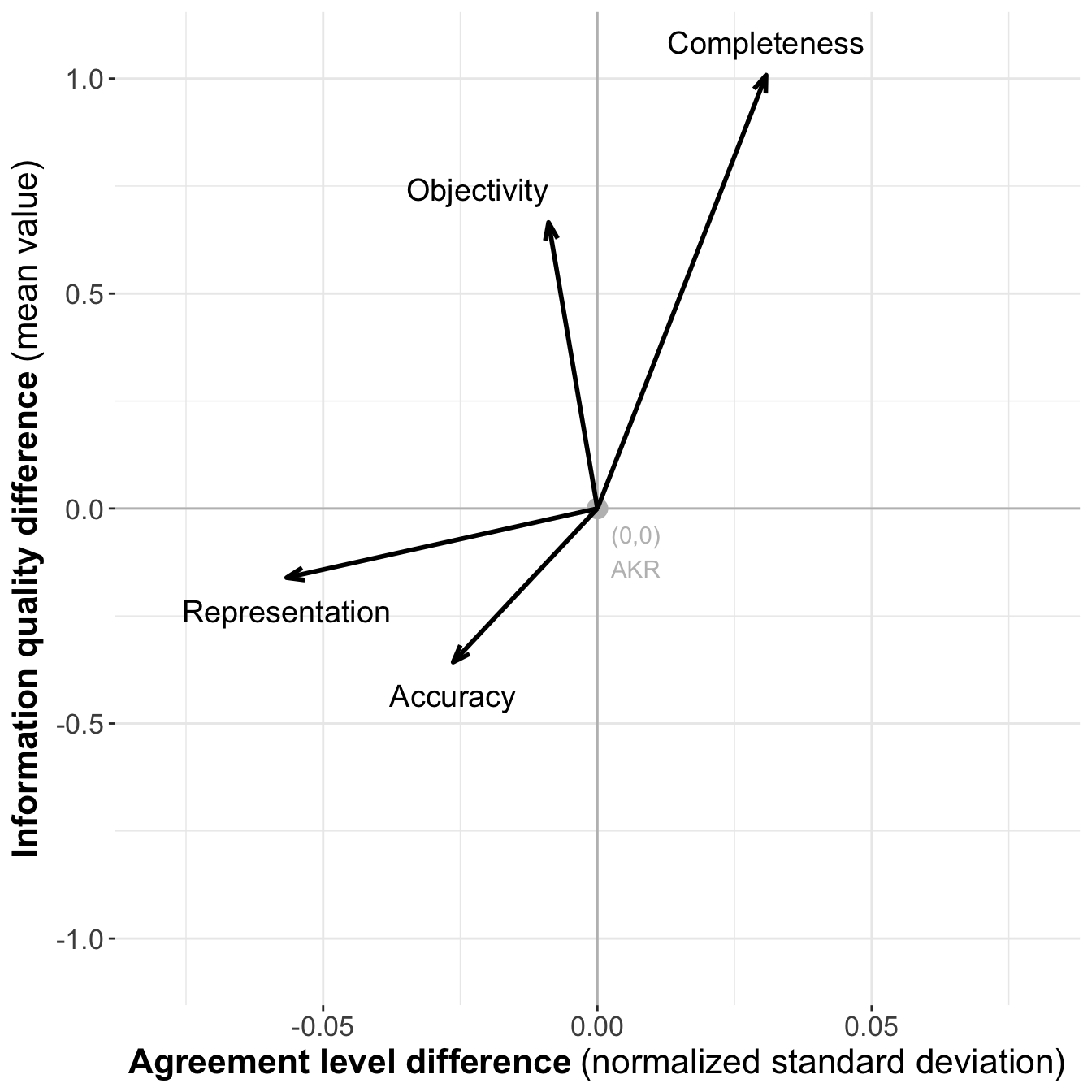}

}

\caption{Difference of agreement level and IQ for individual dimensions as vectors compared to AKR $(0,0)$}\label{fig:IQ-diff-by-dim}
\end{figure}

Impacts per individual IQ dimensions are depicted in Figure
\ref{fig:IQ-diff-by-dim}, where assessment results are compared to the
AKR in even more straight forward manner. The IQ score of completeness
increased the most by \(1,01\) (on \(1 - 7\) scale), followed by
\(0,67\) increase of objectivity, while the IQ score of representation
declined by \(0,16\) and accuracy by \(0,36\). In terms of agreement
level, the normalized standard deviation of completeness slightly
increased by \(0,03\), while it marginally declined for dimensions of
objectivity (by \(0,01\)), accuracy (by \(0,03\)), and Representation
(by \(0,06\)).

\subsection{Discussion of impact on IQ dimensions}\label{discussion}

The IQ score of accuracy decreased due to polarized participant's domain
knowledge distribution across selected articles. Students considered
their domain knowledge about the evaluated article either expert or
ignorant. In all, \(71\%\) of the students rated the article of Occitan
and Alexandrite as \(4\), not being able to agree or disagree regarding
the article's quality. If such assessments are discarded, value for
accuracy actually increases by \(0,77\) points.

Approach of applied Gricean principle which made Wikipedia articles fit
for students needs also increased their perception of quality for
dimensions of objectivity and completeness. Students perceived retrieved
information from modified Wikipedia articles more objective and complete
as students retrieved information from randomly chosen original
Wikipedia articles that were evaluated in the AKR. Thus, the agreement
level between students' scores for completeness increased, making a
dimension more measurable, while agreement level of objectivity stayed
the same.

The quality of objectivity most likely increased because Grice's
conversation maxims and submaxims \citep{grice_logic_1967} were applied
with very succinct writing style that was perceived by students as more
objective as the writing style of Wikipedia editors. In addition, all
given information in re-written articles was cross-checked with other
online sources causing also the perception of accuracy to increase, but
only for students who were confident enough (had enough domain
knowledge) to evaluate this dimension. Both improvements made re-written
articles less biased and more truthful.

Even more noticeable finding is the improvement of completeness. As
discussed in \citep{yaari_information_2011}, article length is not
entirely detached from the content. Authors argue that longer articles,
if written comprehensively, include more information than shorter ones.
In other words it would be difficult to write short articles of good
quality that would still be considered complete. Although application of
Gricean principle made re-written articles only \(180\) words long,
students perceived shorter versions to be more complete, compared to
randomly selected articles from Wikipedia with length from \(200\) to
\(3.500\) words, which were evaluated in the AKR. There are two
implications of this finding.

First, encyclopedias could implement Grice's conversation maxims and
submaxims as editorial guidelines or use proposed structure of
re-written articles as a template for presenting concise summaries of
topics. This study confirmed that there are many possibilities for IQ
improvement and that editing based on Gricean principle would make
articles for students more complete and objective. Further more,
presenting information in more concise manner makes articles fitter for
students since they typically use Wikipedia as a starting point for
further inquiries. Second, researchers could gain more IQ evaluations if
they would shorten their articles. Rater spends only one-fifth of the
time to read an article of \(180\) words as opposed to an article of
\(1.000\) words. Researchers could therefore use each rater to assess
more articles or offer less time demanding surveys.

Despite, providing visually appealing assessment form and consistent
presentation of information in articles (e.g.~same notation for number
and currencies), students did not perceive quality of representation to
improve compared to the AKR. It may be that students are not accustomed
to assessing representation to the same extent as for other IQ
dimensions. In addition, information representation is not so often
critical for daily tasks and even when it becomes critical, it can be
more easily resolved as inaccurate or incomplete information.

\section{Conslusion}\label{conclusion}

This paper proposed and tested an approach of making Wikipedia articles
more relevant for students' needs, thus provided evidence of improved
IQ. \citep{mai_quality_2013} expectations that the pragmatic philosophy
of a language can be employed to make objects more relevant for consumer
needs were confirmed. When Gricean principle was applied on Wikipedia
articles to make them fit for students' needs the average value of
scores for all four IQ dimensions (CIQ) increased by \(0,27\) compared
to the AKR \citep{arazy_measurability_2011}. Students annoyed by low IQ
recognized re-written articles to have significantly higher IQ (for
\(0,18\)) than students who were not annoyed by low IQ. IQ increase was
notable for dimensions of objectivity and completeness, quality of
accuracy decreased due to polarized domain knowledge and representation
obtained similar quality as in the AKR.

This paper provides to authors informed insights on students' perception
of IQ, which can be used as a reference point in future studies. IQ
researchers who employ students as evaluators can validate their
results, whereas non-students-based evaluation studies can compare and
further generalize their findings. Furthermore, results can be
beneficial for Wikipedia and other encyclopedias since they can apply
Gricean principle in a similar way to improve their content for
students' needs. In this manner Wikipedia articles can be shortened,
serving students with information of improved IQ but at the same time
demanding less work for editors. Students can read more comprehensive
articles faster, obtaining information of a higher quality in less time,
while editors would maintain articles of fewer words.

This study has limitations, since it was based on one application of
Grice's conversation maxims and submaxims; on a limited number of
articles; compared to a limited number of recent researches; and
conducted for students. Despite our consulting and academic background,
which promotes succinct written and verbal communication and literature
review on research how and why students use Wikipedia, it is not
conclusive that proposed application of Grice's conversation maxims and
submaxims was the most effective communication of relevant information
to students. IQ of Wikipedia articles can be improved even more. Only
limited number of articles were selected. However, they were selected in
such a manner that students evaluated topics of low and high domain
knowledge. Results were compared only to the AKR, since there is a
deficiency of similar evaluation studies.

Research community is therefore encouraged to conduct similar IQ
improvement studies, not only for students but also for other groups
(e.g.~young professionals, retired people etc.). Proposed approach of
applying Gricean principle can be used as a recommendation on how to
construct articles of fewer words to have shorter questionnaires, obtain
more article evaluations per rater, and ultimately have more reliable
measurements. In the future we hope to repeat our study with a larger
sample size of articles of more diverse raters, based on characteristics
such as age, education, and domain knowledge, so that findings could be
generalized to entire population with higher confidence.


\begin{thebibliography}{}

\bibitem[Arazy and Kopak, 2011]{arazy_measurability_2011}
Arazy, O. and Kopak, R. (2011).
\newblock On the {Measurability} of {Information} {Quality}.
\newblock {\em Journal of the American Society for Information Science and
  Technology}, 62(1):89--99.

\bibitem[Baskarada and Koronios, 2014]{baskarada_critical_2014}
Baskarada, S. and Koronios, A. (2014).
\newblock A {Critical} {Success} {Factor} {Framework} for {Information}
  {Quality} {Management}.
\newblock {\em Information Systems Management}, 31(4):276--295.

\bibitem[Bastida and Huan, 2014]{bastida_performance_2014}
Bastida, U. and Huan, T.~C. (2014).
\newblock Performance evaluation of tourism websites' information quality of
  four global destination brands: {Beijing}, {Hong} {Kong}, {Shanghai}, and
  {Taipei}.
\newblock {\em Journal of Business Research}, 67(2):167--170.

\bibitem[Bates, 2005a]{bates_information_2005}
Bates, M.~J. (2005a).
\newblock Information and knowledge: an evolutionary framework for information
  science.
\newblock {\em Information Research}, 10(4).

\bibitem[Bates, 2005b]{bates_introduction_2005}
Bates, M.~J. (2005b).
\newblock An {Introduction} to {Metatheories}, {Theories}, and {Models}.
\newblock In {\em Theories of {Information} {Behaviour}}, pages 1--24.

\bibitem[Berezan et~al., 2016]{berezan_impact_2016}
Berezan, O., Yoo, M., and Christodoulidou, N. (2016).
\newblock The impact of communication channels on communication style and
  information quality for hotel loyalty programs.
\newblock {\em Journal of Hospitality and Tourism Technology}, 7(1):100--116.

\bibitem[Brown, 2011]{brown_wikipedia_2011}
Brown, A.~R. (2011).
\newblock Wikipedia as a {Data} {Source} for {Political} {Scientists}:
  {Accuracy} and {Completeness} of {Coverage}.
\newblock {\em PS - Political Science \& Politics}, 44(2):339--343.

\bibitem[Ceylan et~al., 2016]{ceylan_information_2016}
Ceylan, H.~H., Caypinar, B., Kucukkoc, M., Uzuner, S., and Kucukdurmaz, F.
  (2016).
\newblock Information {Quality} on {Developmental} {Dysplasia} of the {Hip} on
  {Turkish} {Websites}.
\newblock {\em Journal of Academic Research in Medicine-Jarem}, 6(2):84--87.

\bibitem[Chesney, 2006]{chesney_empirical_2006}
Chesney, T. (2006).
\newblock An empirical examination of {Wikipedia}'s credibility.
\newblock {\em First monday}, 11(11).

\bibitem[Choolhun, 2009]{choolhun_google:_2009}
Choolhun, N. (2009).
\newblock Google: to use, or not to use. {What} is the question?
\newblock {\em Legal Information Management}, 9(3):168--172.

\bibitem[Clauson et~al., 2008]{clauson_scope_2008}
Clauson, K.~A., Polen, H.~H., Boulos, M. N.~K., and Dzenowagis, J.~H. (2008).
\newblock Scope, {Completeness}, and {Accuracy} of {Drug} {Information} in
  {Wikipedia}.
\newblock {\em Annals of Pharmacotherapy}, 42(12):1814--1821.

\bibitem[Corona et~al., 2015]{corona_accounting_2015}
Corona, C., Nan, L., and Zhang, G.~Q. (2015).
\newblock Accounting {Information} {Quality}, {Interbank} {Competition}, and
  {Bank} {Risk}-{Taking}.
\newblock {\em Accounting Review}, 90(3):967--985.

\bibitem[Denning et~al., 2005]{denning_wikipedia_2005}
Denning, P., Horning, J., Parnas, D., and Weinstein, L. (2005).
\newblock Wikipedia risks.
\newblock {\em Communications of the ACM}, 48(12):152--152.

\bibitem[Detlor et~al., 2013]{detlor_information_2013}
Detlor, B., Hupfer, M.~E., Ruhi, U., and Zhao, L. (2013).
\newblock Information quality and community municipal portal use.
\newblock {\em Government Information Quarterly}, 30(1):23--32.

\bibitem[Eppler, 2006]{eppler_managing_2006}
Eppler, M.~J. (2006).
\newblock {\em Managing {Information} {Quality}: {Increasing} the {Value} of
  {Information} in {Knowledge}-intensive {Products} and {Processes}}.
\newblock Springer-Verlag, Berlin, Germany, 2nd edition edition.

\bibitem[Fallis, 2008]{fallis_toward_2008}
Fallis, D. (2008).
\newblock Toward an epistemology of {Wikipedia}.
\newblock {\em Journal of the American Society for Information Science and
  Technology}, 59(10):1662--1674.

\bibitem[Fidler and Lavbič, 2015]{fidler_research_2015}
Fidler, M. and Lavbič, D. (2015).
\newblock Research {About} {Measurability} of {Information} {Quality}.
\newblock In {\em 10th {International} {Conference} on {Knowledge} {Management}
  in {Organisations} ({KMO} 2015)}, Slovenia, Maribor.

\bibitem[Fink-Shamit and Bar-Ilan, 2008]{fink-shamit_information_2008}
Fink-Shamit, N. and Bar-Ilan, J. (2008).
\newblock Information quality assessment on the {Web} - an expression of
  behaviour.
\newblock {\em Information Research - an International Electronic Journal},
  13(4).

\bibitem[Font et~al., 2015]{font_assessing_2015}
Font, L., Zouaq, A., and Gagnon, M. (2015).
\newblock {\em Assessing the {Quality} of {Domain} {Concepts} {Descriptions} in
  {DBpedia}}.
\newblock 2015 11th {International} {Conference} on {Signal}-{Image}
  {Technology} \& {Internet}-{Based} {Systems}. Ieee, New York.

\bibitem[Ge and Helfert, 2007]{ge_review_2007}
Ge, M. and Helfert, M. (2007).
\newblock {\em A {Review} of {Information} {Quality} {Research} - {Develop} a
  {Research} {Agenda}}.

\bibitem[Ghasemaghaei and Hassanein, 2016]{ghasemaghaei_macro_2016}
Ghasemaghaei, M. and Hassanein, K. (2016).
\newblock A macro model of online information quality perceptions: {A} review
  and synthesis of the literature.
\newblock {\em Computers in Human Behavior}, 55:972--991.

\bibitem[Grice, 1967]{grice_logic_1967}
Grice, H.~P. (1967).
\newblock {\em Logic and {Conversation}}.
\newblock Harvard University Press.

\bibitem[Haakonsen~Dahl and Jørgensen, 2014]{haakonsen_dahl_intra-_2014}
Haakonsen~Dahl, S.~S. and Jørgensen, L. (2014).
\newblock Intra- and {Inter}-{Rater} {Reliability} of the {Mini}-{Balance}
  {Evaluation} {Systems} {Test} in {Individuals} with {Stroke}.
\newblock {\em International Journal of Physical Medicine \& Rehabilitation},
  2(1).

\bibitem[Helfert, 2001]{helfert_managing_2001}
Helfert, M. (2001).
\newblock {\em Managing and {Measuring} {Data} {Quality} in {Data}
  {Warehousing}}.

\bibitem[Helfert et~al., 2013]{helfert_impact_2013}
Helfert, M., Walshe, R., and Gurrin, C. (2013).
\newblock The {Impact} of {Information} {Quality} on {Quality} of {Life}: {An}
  {Information} {Quality} {Oriented} {Framework}.
\newblock {\em Ieice Transactions on Communications}, E96B(2):404--409.

\bibitem[Hilligoss and Rieh, 2008]{hilligoss_developing_2008}
Hilligoss, B. and Rieh, S.~Y. (2008).
\newblock Developing a unifying framework of credibility assessment:
  {Construct}, heuristics, and interaction in context.
\newblock {\em Information Processing \& Management}, 44(4):1467--1484.

\bibitem[Hjørland, 2007]{hjorland_information:_2007}
Hjørland, B. (2007).
\newblock Information: {Objective} or subjective/situational?
\newblock {\em Journal of the American Society for Information Science and
  Technology}, 58(10):1448--1456.

\bibitem[Kahn et~al., 2002]{kahn_information_2002}
Kahn, B.~K., Strong, D.~M., and Wang, R.~Y. (2002).
\newblock Information {Quality} {Benchmarks}: {Product} and {Service}
  {Performance}.
\newblock {\em Communications of the ACM}, 45(4).

\bibitem[Kambatla et~al., 2014]{kambatla_trends_2014}
Kambatla, K., Kollias, G., Kumar, V., and Grama, A. (2014).
\newblock Trends in big data analytics.
\newblock {\em Journal of Parallel and Distributed Computing},
  74(7):2561--2573.

\bibitem[Knight and Burn, 2005]{knight_developing_2005}
Knight, S.-A. and Burn, J. (2005).
\newblock Developing a {Framework} for {Assessing} {Information} {Quality} on
  the {World} {Wide} {Web}.
\newblock {\em Informing Science Journal}, 8:159--172.

\bibitem[Kogut and Zander, 1996]{kogut_what_1996}
Kogut, B. and Zander, U. (1996).
\newblock What firms do? {Coordination}, identity, and learning.
\newblock {\em Organization Science}, 7(5):502--518.

\bibitem[Korosec et~al., 2010]{korosec_chemical_2010}
Korosec, L., Limacher, P.~A., Lüthi, H.~P., and Brändle, M.~P. (2010).
\newblock Chemical information media in the chemistry lecture hall: a
  comparative assessment of two online encyclopedias.
\newblock {\em CHIMIA International Journal for Chemistry}, 64(5):309--314.

\bibitem[Lee et~al., 2016]{lee_enhancing_2016}
Lee, R.~P., Chen, Q.~M., and Hartmann, N.~N. (2016).
\newblock Enhancing {Stock} {Market} {Return} with {New} {Product}
  {Preannouncements}: {The} {Role} of {Information} {Quality} and
  {Innovativeness}.
\newblock {\em Journal of Product Innovation Management}, 33(4):455--471.

\bibitem[Leite et~al., 2016]{leite_model_2016}
Leite, P., Goncalves, J., Teixeira, P., and Rocha, A. (2016).
\newblock A model for the evaluation of data quality in health unit websites.
\newblock {\em Health Informatics Journal}, 22(3):479--495.

\bibitem[Lim, 2009]{lim_how_2009}
Lim, S. (2009).
\newblock How and {Why} {Do} {College} {Students} {Use} {Wikipedia}?
\newblock {\em Journal of the American Society for Information Science and
  Technology}, 60(11):2189--2202.

\bibitem[Lim and Kwon, 2010]{lim_gender_2010}
Lim, S. and Kwon, N. (2010).
\newblock Gender differences in information behavior concerning {Wikipedia}, an
  unorthodox information source?
\newblock {\em Library \& Information Science Research}, 32(3):212--220.

\bibitem[Lopez et~al., 2016]{lopez_information_2016}
Lopez, D.~M., Blobel, B., and Gonzalez, C. (2016).
\newblock Information quality in healthcare social media - an architectural
  approach.
\newblock {\em Health and Technology}, 6(1):17--25.

\bibitem[Luyt et~al., 2008]{luyt_improving_2008}
Luyt, B., Aaron, T. C.~H., Thian, L.~H., and Hong, C.~K. (2008).
\newblock Improving {Wikipedia}'s accuracy: {Is} edit age a solution?
\newblock {\em Journal of the American Society for Information Science and
  Technology}, 59(2):318--330.

\bibitem[Madhikermi et~al., 2016]{madhikermi_data_2016}
Madhikermi, M., Kubler, S., Robert, J., Buda, A., and Framling, K. (2016).
\newblock Data quality assessment of maintenance reporting procedures.
\newblock {\em Expert Systems with Applications}, 63:145--164.

\bibitem[Mai, 2013]{mai_quality_2013}
Mai, J.~E. (2013).
\newblock The quality and qualities of information.
\newblock {\em Journal of the American Society for Information Science and
  Technology}, 64(4):675--688.

\bibitem[Mesgari et~al., 2015]{mesgari_sum_2015}
Mesgari, M., Okoli, C., Mehdi, M., Nielsen, F.~A., and Lanamaki, A. (2015).
\newblock "{The} {Sum} of {All} {Human} {Knowledge}": {A} {Systematic} {Review}
  of {Scholarly} {Research} on the {Content} of {Wikipedia}.
\newblock {\em Journal of the Association for Information Science and
  Technology}, 66(2):219--245.

\bibitem[Metzger, 2007]{metzger_making_2007}
Metzger, M.~J. (2007).
\newblock Making sense of credibility on the web: {Models} for evaluating
  online information and recommendations for future research.
\newblock {\em Journal of the American Society for Information Science and
  Technology}, 58(13):2078--2091.

\bibitem[Michnik and Lo, 2009]{michnik_assessment_2009}
Michnik, J. and Lo, M.~C. (2009).
\newblock The assessment of the information quality with the aid of multiple
  criteria analysis.
\newblock {\em European Journal of Operational Research}, 195(3):850--856.

\bibitem[Moe-Nilssen et~al., 2008]{moe-nilssen_criteria_2008}
Moe-Nilssen, R., Nordin, E., and Lundin-Olsson, L. (2008).
\newblock Criteria for evaluation of measurement properties of clinical balance
  measures for use in fall prevention studies.
\newblock {\em Journal of Evaluation in Clinical Practice}, 14(2):236--240.

\bibitem[Mohammadi et~al., 2015]{mohammadi_is_2015}
Mohammadi, F., Abrizah, A., and Nazari, M. (2015).
\newblock Is the information fit for use? {Exploring} teachers perceived
  information quality indicators for {Farsi} web-based learning resources.
\newblock {\em Malaysian Journal of Library \& Information Science},
  20(1):99--122.

\bibitem[Nicolaou et~al., 2013]{nicolaou_information_2013}
Nicolaou, A.~I., Ibrahim, M., and van Heck, E. (2013).
\newblock Information quality, trust, and risk perceptions in electronic data
  exchanges.
\newblock {\em Decision Support Systems}, 54(2):986--996.

\bibitem[Okoli et~al., 2014]{okoli_wikipedia_2014}
Okoli, C., Mesgari, M., Mehdi, M., Nielsen, F.~A., and Lanamaki, A. (2014).
\newblock Wikipedia in the {Eyes} of {Its} {Beholders}: {A} {Systematic}
  {Review} of {Scholarly} {Research} on {Wikipedia} {Readers} and {Readership}.
\newblock {\em Journal of the Association for Information Science and
  Technology}, 65(12):2381--2403.

\bibitem[Paglieri et~al., 2014]{paglieri_trusting_2014}
Paglieri, F., Castelfranchi, C., Pereira, C.~D., Falcone, R., Tettamanzi, A.,
  and Villata, S. (2014).
\newblock Trusting the messenger because of the message: feedback dynamics from
  information quality to source evaluation.
\newblock {\em Computational and Mathematical Organization Theory},
  20(2):176--194.

\bibitem[Pannucci and Wilkins, 2010]{pannucci_identifying_2010}
Pannucci, C.~J. and Wilkins, E.~G. (2010).
\newblock Identifying and {Avoiding} {Bias} in {Research}.
\newblock {\em Plastic and Reconstructive Surgery}, 126(2):619--625.

\bibitem[Patch, 2010]{patch_meeting_2010}
Patch, P. (2010).
\newblock Meeting {Student} {Writers} {Where} {They} {Are}: {Using} {Wikipedia}
  to {Teach} {Responsible} {Scholarship}.
\newblock {\em Teaching English in the Two-Year College}, 37(3):278--285.

\bibitem[Petter et~al., 2013]{petter_information_2013}
Petter, S., delone, W., and McLean, E.~R. (2013).
\newblock Information {Systems} {Success}: {The} {Quest} for the {Independent}
  {Variables}.
\newblock {\em Journal of Management Information Systems}, 29(4):7--61.

\bibitem[Rieh and Danielson, 2007]{rieh_credibility:_2007}
Rieh, S.~Y. and Danielson, D.~R. (2007).
\newblock Credibility: {A} multidisciplinary framework.
\newblock {\em Annual Review of Information Science and Technology},
  41:307--364.

\bibitem[Romero-Rodriguez et~al., 2016]{romero-rodriguez_dimensions_2016}
Romero-Rodriguez, L.~M., de~Casas-Moreno, P., and Torres-Toukoumidis, A.
  (2016).
\newblock Dimensions and {Indicators} of the {Information} {Quality} in
  {Digital} {Media}.
\newblock {\em Comunicar}, (49):91--100.

\bibitem[Sackmann, 1991]{sackmann_cultural_1991}
Sackmann, S.~A. (1991).
\newblock {\em Cultural {Knowledge} in {Organizations}: {Exploring} the
  {Collective} {Mind}}.
\newblock SAGE Publications.

\bibitem[Savchenko, 2003]{savchenko_automating_2003}
Savchenko, S. (2003).
\newblock Automating {Objective} {Data} {Quality} {Assessment} ({Experiences}
  in {Software} {Tool} {Design}).
\newblock In {\em International {Conference} on {Information} {Quality} ({IQ}
  2003)}.

\bibitem[Shaw, 2008]{shaw_wikipedia_2008}
Shaw, D. (2008).
\newblock Wikipedia in the {Newsroom}.
\newblock {\em American Journalism Review}, pages 40--45.

\bibitem[Shen et~al., 2013]{shen_unleash_2013}
Shen, X.~L., Wang, N., Sun, Y.~Q., and Xiang, L. (2013).
\newblock Unleash the power of mobile word-of-mouth {An} empirical study of
  system and information characteristics in ubiquitous decision making.
\newblock {\em Online Information Review}, 37(1):42--60.

\bibitem[Stenmark, 2000]{stenmark_leveraging_2000}
Stenmark, D. (2000).
\newblock Leveraging tacit organizational knowledge.
\newblock {\em Journal of Management Information Systems}, 17(3):9--24.

\bibitem[Straub et~al., 2004]{straub_validation_2004}
Straub, D., Boudreau, M.-C., and Gefen, D. (2004).
\newblock Validation guidelines for {IS} positivist research.
\newblock {\em The Communications of the Association for Information Systems},
  13(1):380--427.

\bibitem[Stvilia et~al., 2008]{stvilia_information_2008}
Stvilia, B., Twidale, M.~B., Smith, L.~C., and Gasser, L. (2008).
\newblock Information quality work organization in {Wikipedia}.
\newblock {\em Journal of the American Society for Information Science and
  Technology}, 59(6):983--1001.

\bibitem[{Suhardi} et~al., 2014]{suhardi_total_2014}
{Suhardi}, Gunawan, G., Dewi, A.~Y., and {Ieee} (2014).
\newblock Total {Information} {Quality} {Management}-{Capability} {Maturity}
  {Model} ({TIQM}-{CMM}): {An} {Information} {Quality} {Management} {Maturity}
  {Model}.
\newblock {\em 2014 International Conference on Data and Software Engineering
  (ICODSE)}, page~6.

\bibitem[Vrhovec et~al., 2015]{vrhovec_outsourcing_2015}
Vrhovec, S. L.~R., Trkman, M., Kumer, A., Krisper, M., and Vavpotic, D. (2015).
\newblock Outsourcing as an {Economic} {Development} {Tool} in {Transition}
  {Economies}: {Scattered} {Global} {Software} {Development}.
\newblock {\em Information Technology for Development}, 21(3):445--459.

\bibitem[Wallace and Van~Fleet, 2005]{wallace_democratization_2005}
Wallace, D.~P. and Van~Fleet, C. (2005).
\newblock The democratization of information? {Wikipedia} as a reference
  resource.
\newblock {\em Reference \& User Services Quarterly}, 45(2):100--103.

\bibitem[Wand and Wang, 1996]{wand_anchoring_1996}
Wand, Y. and Wang, R.~Y. (1996).
\newblock Anchoring data quality dimensions in ontological foundations.
\newblock {\em Communications of the ACM}, 39(11):86--95.

\bibitem[Wang and Strong, 1996]{wang_beyond_1996}
Wang, R.~Y. and Strong, D.~M. (1996).
\newblock Beyond accuracy: what data quality means to data consumers.
\newblock {\em Journal of Management Information Systems}, 12(4):5--33.

\bibitem[Wang et~al., 2014]{wang_model_2014}
Wang, Y.~S., Li, H.~T., Li, C.~R., and Wang, C. (2014).
\newblock A model for assessing blog-based learning systems success.
\newblock {\em Online Information Review}, 38(7):969--990.

\bibitem[Waters, 2007]{waters_why_2007}
Waters, N.~L. (2007).
\newblock Why you can't cite {Wikipedia} in my class.
\newblock {\em Communications of the ACM}, 50(9):15--17.

\bibitem[Yaari et~al., 2011]{yaari_information_2011}
Yaari, E., Baruchson-Arbib, S., and Bar-Ilan, J. (2011).
\newblock Information quality assessment of community-generated content - {A}
  user study of {Wikipedia}.
\newblock {\em Journal of Information Science}, 37(5):487--498.

\bibitem[Zhang et~al., 2005]{zhang_domain_2005}
Zhang, X.~M., Anghelescu, H. G.~B., and Yuan, X.~J. (2005).
\newblock Domain knowledge, search behaviour, and search effectiveness of
  engineering and science students: an exploratory study.
\newblock {\em Information Research - An International Electronic Journal},
  10(2).

\bibitem[Zheng et~al., 2013]{zheng_impacts_2013}
Zheng, Y.~M., Zhao, K.~X., and Stylianou, A. (2013).
\newblock The impacts of information quality and system quality on users'
  continuance intention in information-exchange virtual communities: {An}
  empirical investigation.
\newblock {\em Decision Support Systems}, 56:513--524.

\bibitem[Zhou et~al., 2014]{zhou_supply_2014}
Zhou, H.~G., Shou, Y.~Y., Zhai, X., Li, L., Wood, C., and Wu, X.~B. (2014).
\newblock Supply chain practice and information quality: {A} supply chain
  strategy study.
\newblock {\em International Journal of Production Economics}, 147:624--633.

\end{thebibliography}
\end{document}